\documentclass[sigconf,natbib=true,anonymous=false]{acmart}
\AtBeginDocument{%
  }

\usepackage{multirow}
\usepackage{booktabs}
\usepackage{float}

\copyrightyear{2026}
\acmYear{2026}
\setcopyright{cc}
\setcctype{by}
\acmConference[SIGIR '26]{2026 ACM SIGIR Conference on Research and Development in Information Retrieval}{July 20-24, 2026}{Melbourne, Australia}
\acmBooktitle{2026 ACM SIGIR Conference on Research and Development in Information Retrieval (SIGIR '26), July 20-24, 2026, Melbourne, Australia}
\acmPrice{}
\acmDOI{xxx}
\acmISBN{xxx}
\begin{document}

\title{ECHO: An Open Research Platform for Evaluation of Chat, Human Behavior, and Outcomes}


\author{Jiqun Liu}
\orcid{0000-0003-3643-2182}
\affiliation{%
  \institution{The University of Oklahoma}
  \streetaddress{401 W Brooks Street}
  \city{Norman}
 \state{OK}
  \country{USA}}
\email{jiqunliu@ou.edu}

\author{Nischal Dinesh}
\affiliation{%
  \institution{The University of Oklahoma}
  \streetaddress{401 W Brooks Street}
  \city{Norman}
 \state{OK}
  \country{USA}}
\email{nischal@ou.edu}

\author{Ran Yu}
\affiliation{%
  \institution{GESIS – Leibniz Institute for the Social Sciences}
  \city{Mannheim}
  \country{Germany}}
\email{ran.yu@gesis.org}

\renewcommand{\shortauthors}{He et al.}

\begin{abstract}
ECHO (\textbf{E}valuation of \textbf{C}hat, \textbf{H}uman behavior, and \textbf{O}utcomes) is an open research platform designed to support reproducible, mixed-method studies of human interaction with both conversational AI systems and Web search engines. It enables researchers from varying disciplines to orchestrate end-to-end experimental workflows that integrate consent and background surveys, chat-based and search-based information-seeking sessions, writing or judgment tasks, and pre- and post-task evaluations within a \textit{unified, low-coding-load} framework. ECHO logs fine-grained interaction traces and participant responses, and exports structured datasets for downstream analysis. By supporting both chat and search alongside flexible evaluation instruments, ECHO lowers technical barriers for studying learning, decision making, and user experience across different information access paradigms, empowering researchers from information retrieval, HCI, and the social sciences to conduct scalable and reproducible human-centered AI evaluations. The demo video recording of ECHO platform can be accessed here~\footnote{ \url{https://drive.google.com/file/d/1T16fFcsGkQIPIIHZsaMCFw8yEAn2V5ER/view}}.
\end{abstract}

\ccsdesc[500]{Information systems~Search interfaces}
\ccsdesc[500]{Human-centered computing~User studies}

\keywords{conversational search, large language models, source citation, credibility, transparency, information seeking,  attitude change}


\maketitle

\section{Introduction}
The rapid adoption of large language model (LLM)–based conversational systems is reshaping how users seek, interpret, and act on information. Unlike traditional ranked retrieval systems, conversational interfaces support multi-turn interactions that blend information access with explanation, synthesis, and guidance, often influencing learning, trust, and decisions~\cite{zamani2023conversational}. As a result, conversational and generative systems challenge long-standing assumptions in information retrieval (IR) evaluation, which have historically emphasized static relevance judgments, ranking effectiveness, and efficiency metrics~\cite{carterette2016evaluating, voorhees2005trec}. While these system-oriented measures remain essential, they are increasingly insufficient for capturing how users experience conversational systems over time, how expectations evolve across turns, or how interaction dynamics shape outcomes such as understanding, confidence, and behavioral change~\cite{kelly2009methods}. Recent work has further shown that conversational and generative search systems can systematically influence user judgment through framing, credibility cues, and interaction patterns that are not observable through offline evaluation alone~\cite{mo2025survey, wan2023biasasker, liu2025report}.

Despite growing recognition of these challenges, existing evaluation practices remain fragmented across research communities. System-oriented IR research continues to rely primarily on offline benchmarks and query-based relevance metrics~\cite{chapelle2009expected, meng2026re}, while human-centered studies often employ experimental setups focused on user perceptions, attitudes, tasks or decision-making~\cite{shah2023taking}. In conversational IR, this divide is particularly pronounced: studies examining dialogue quality, trust, or satisfaction are frequently disconnected from fine-grained system instrumentation, and comparisons between AI agents and traditional Web search are often conducted using separate tools and incompatible workflows~\cite{lei2020conversational, liu2022toward, liu2019interactive}. This fragmentation makes it difficult to connect user-level phenomena, such as cognitive bias, expectation mismatch, or strategy shifts~\cite{azzopardi2021cognitive, liu2023toward}, to underlying system behavior, and hinders cumulative knowledge building across IR, HCI, cognitive and behavioral sciences~\cite{liu2022toward}. 

Existing experimental systems are poorly suited for studying conversational and generative information access in an integrated way. Classical IR infrastructures based on Cranfield-style test collections and TREC primarily support offline, query-level relevance evaluation and offer limited visibility into interaction dynamics or user outcomes over time~\cite{carterette2016evaluating}. Although interactive IR research emphasizes user-centered evaluation, most studies rely on bespoke experimental interfaces that are difficult to reuse or adapt across settings~\cite{kelly2009methods, jiang2025landscape}. Recent search and LLM evaluations typically employ standalone chat tools or platform-specific logs, which lack unified workflow control and longitudinal outcome measures~\cite{lei2020conversational, zamani2023conversational, white2025information}.

To address this challenge, we introduce \textbf{ECHO} (Evaluation of Chat, Human behavior, and Outcomes), an open research platform designed to support \textit{end-to-end, mixed-method evaluation} of both conversational AI systems and Web search. \textbf{ECHO enables researchers to configure complete experimental workflows through a flexible administrative interface rather than code}. By automatically logging fine-grained interaction traces alongside participant responses, ECHO facilitates studies that jointly examine system behavior, human interaction, and downstream outcomes. ECHO is designed as shared evaluation infrastructure rather than a task-specific experimental artifact, lowering technical barriers for researchers across varying fields to conduct replicable studies. 


\section{System Overview}

This section describes the user interface and system architecture of ECHO. At a high level, ECHO enables researchers to configure experimental workflows through an administrator dashboard, while participants complete consent forms, surveys, and information-seeking tasks through a web-based interface. The platform supports two primary task modalities: a ChatGPT style conversational interface powered by LLM APIs, and a web search interface powered by the Search API. All participant interactions, including prompts, responses, queries, clicks, and survey answers, are automatically logged to Firebase and can be exported as structured CSV files.

\subsection{Deployment and Resources Requirement}
ECHO is designed to be deployed and used by individual researchers or research teams with minimal technical overhead or coding effort. To run the system, researchers set up a local or institutional instance of the web application by installing standard dependencies (Node.js and a modern Web browser) and connecting the application to a Firebase project using the project-provided configuration files. No dedicated server, cluster, or specialized hardware is required, as all backend functionality, including authentication, data storage, and logging, is handled through Firebase’s serverless infrastructure, and model inference is accessed via external LLM and search APIs. Study configuration, task design, and survey construction are performed entirely through the administrator dashboard, allowing researchers to define experimental workflows, instruments, and triggers without writing custom code. As a result, deploying ECHO primarily involves specifying API credentials and study parameters rather than implementing interfaces or logging pipelines. This lightweight setup enables researchers, particularly those from IR-adjacent fields, HCI, social sciences, and humanities, to locally host and operate the system with modest computational resources, while maintaining full control over experimental design, data collection, and export for downstream analysis.

\begin{table}[h]
\small
\caption{System Prerequisites}
\label{tab:prerequisites}
\begin{tabular}{@{}ll@{}}
\toprule
Requirement & Description \\
\midrule
Node.js (v14+) & JavaScript runtime environment \\
npm & Package manager (included with Node.js) \\
Firebase account & Free tier sufficient for most studies \\
LLM API key & OpenAI, Gemini, or Claude API access \\
Search API key &  Search API access \\
\bottomrule
\end{tabular}
\end{table}

To deploy ECHO, researchers need to clone the repository and install dependencies using standard Node.js tooling:

\begin{verbatim}
git clone https://github.com/OUHCIRGroup/echo
cd echo && npm install
\end{verbatim}

Next, researchers can create a Firebase project and configure the application by copying the provided environment template (\texttt{cp .env.example .env}) and adding their Firebase credentials (API key, project ID, authentication domain). After starting the application with \texttt{npm start}, researchers access the admin setup page to create an administrator account and configure LLM and search API keys through the dashboard.

\subsection{Participant Interface}

The participant facing interface guides users through a configurable study workflow. Upon authentication, participants encounter:

\textbf{Consent and Background Surveys.} Participants first review an IRB-compliant consent form with customizable text and acknowledgment checkboxes. Following consent, a demographic questionnaire collects background information using configurable question types (Likert scales, multiple choice, open-ended responses).

\begin{figure}[h]
  \centering
  \includegraphics[width=\columnwidth]{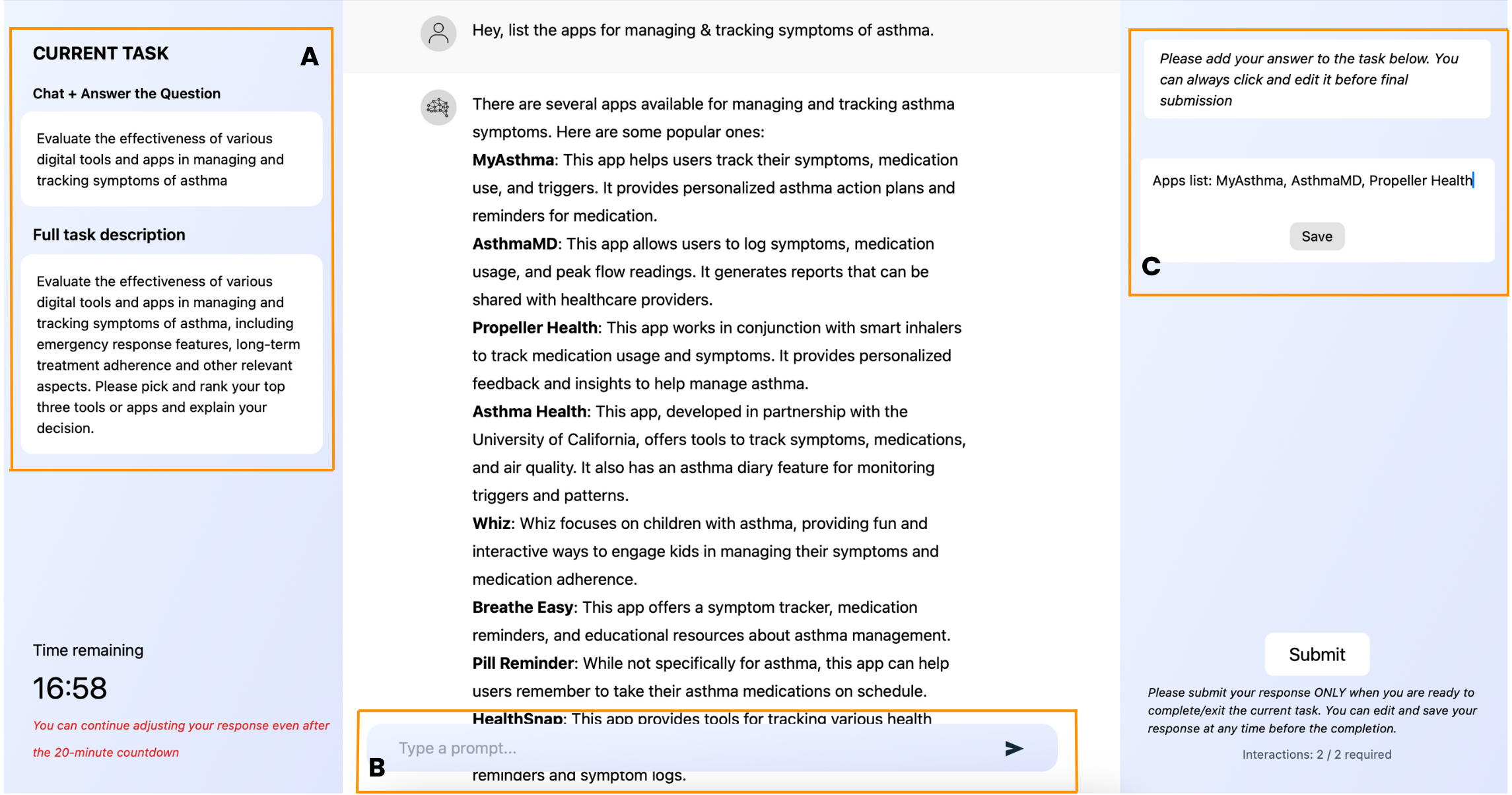}
  \caption{The ECHO participant interaction interface. (a) Task description panel displaying current research task. (b) Chat conversation area with text input for prompts. (c) Optional note taking sidebar for participant responses.}
  \label{fig:interface}
\end{figure}

\textbf{Chat Task Interface.} The chat interface  presents a task description panel on the left (Figure~\ref{fig:interface}a), a central conversation area supporting markdown rendering and code highlighting, and an optional note-taking sidebar on the right (Figure~\ref{fig:interface}c). Users submit prompts through a text input field, and LLM responses stream in real-time. An \textit{in-situ} response rating mechanism allows participants to assess response quality during chat interactions.

\textbf{Search Task Interface.} The search interface  mirrors the chat layout but replaces the conversation area with a search box and results display. Search results show titles, URLs, and abstracts in snippets on SERPs. Clickthrough behavior is logged, and the system captures result ranking for clicked Web pages.

\textbf{Pre/Post-Task Questionnaires.} Before and after each task, participants complete questionnaires assessing expectations and fulfillment using a configurable intention typology framework. Attention check questions can be embedded to ensure response quality.

\subsection{Administrator Dashboard}
The administrator dashboard enables researchers to configure and deploy customized user studies without coding. As shown in Table~\ref{tab:admin}, the dashboard provides centralized control over study settings, workflow management, survey instruments, API configuration, and data export. Survey instruments support both Form and JSON editing modes for flexible configuration. The \textit{In-Situ Surveys} feature allows researchers to configure popup questionnaires triggered by participant actions. All collected data can be exported as structured CSV files through \textit{View Responses}.

\begin{table}[h]
\small
\caption{Administrator Dashboard Features}
\label{tab:admin}
\begin{tabular}{@{}ll@{}}
\toprule
Feature & Description \\
\midrule
Study Settings & Task order, notes toggle, minimum interactions \\
Manage Study Flow & Enable/disable and reorder workflow steps \\
Insert/View Tasks & Create research task descriptions \\
Manage Surveys & Edit demographic and experience questions \\
Manage Typology & Configure intention classification categories \\
API Settings & Manage LLM API and search API credentials \\
In-Situ Surveys & Configure popup surveys triggered by actions \\
View Responses & Browse and export participant data \\
\bottomrule
\end{tabular}
\end{table}

\begin{figure}[h]
  \centering
  \includegraphics[width=\columnwidth]{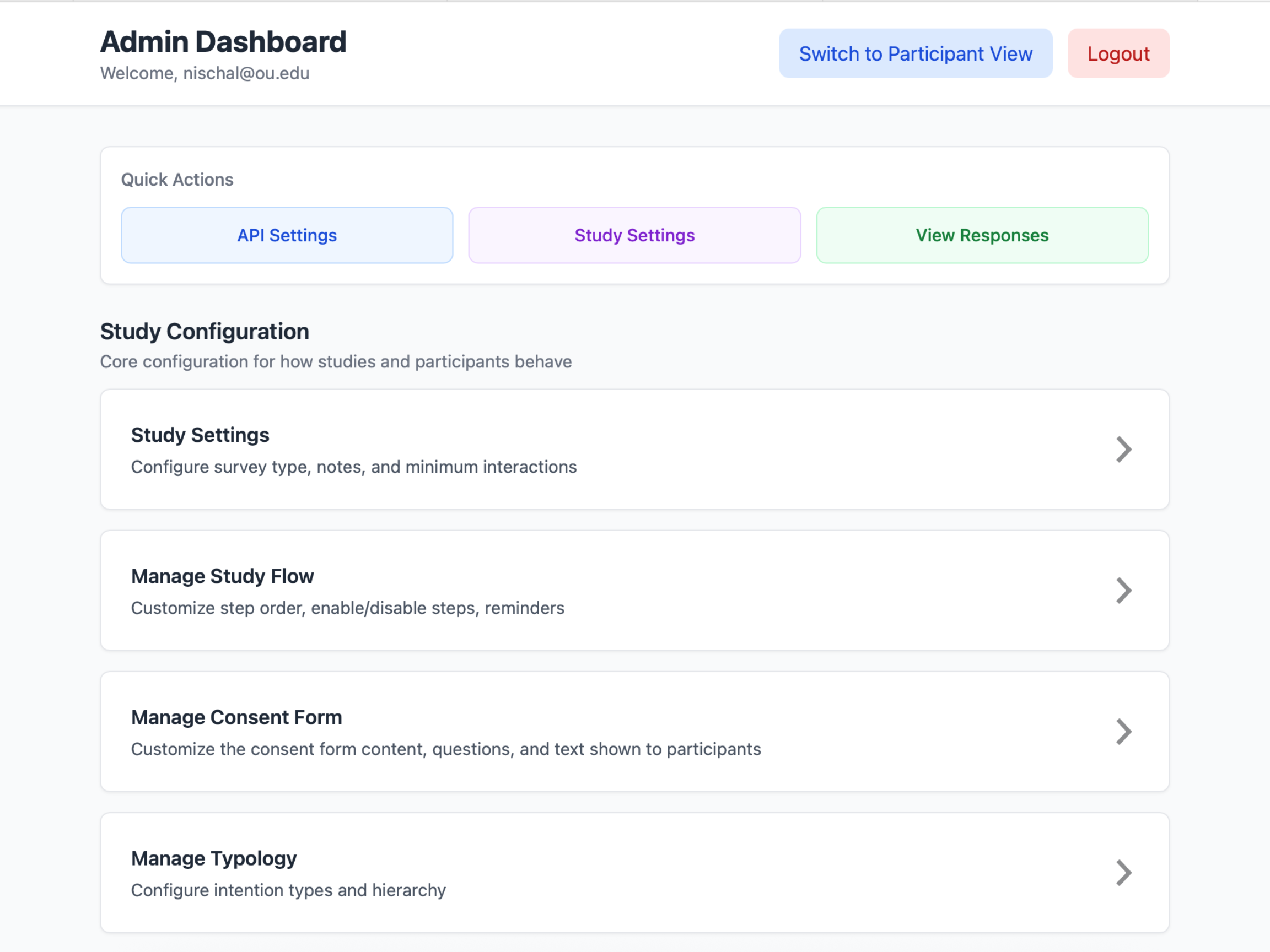}
  \caption{ECHO administrator dashboard for study configuration and management. The study components can be customized through the admin interface rather than code.}
  \label{fig:admin-interface}
\end{figure}

\subsection{System Architecture}

\begin{figure}[h]
  \centering
  \includegraphics[width=\columnwidth]{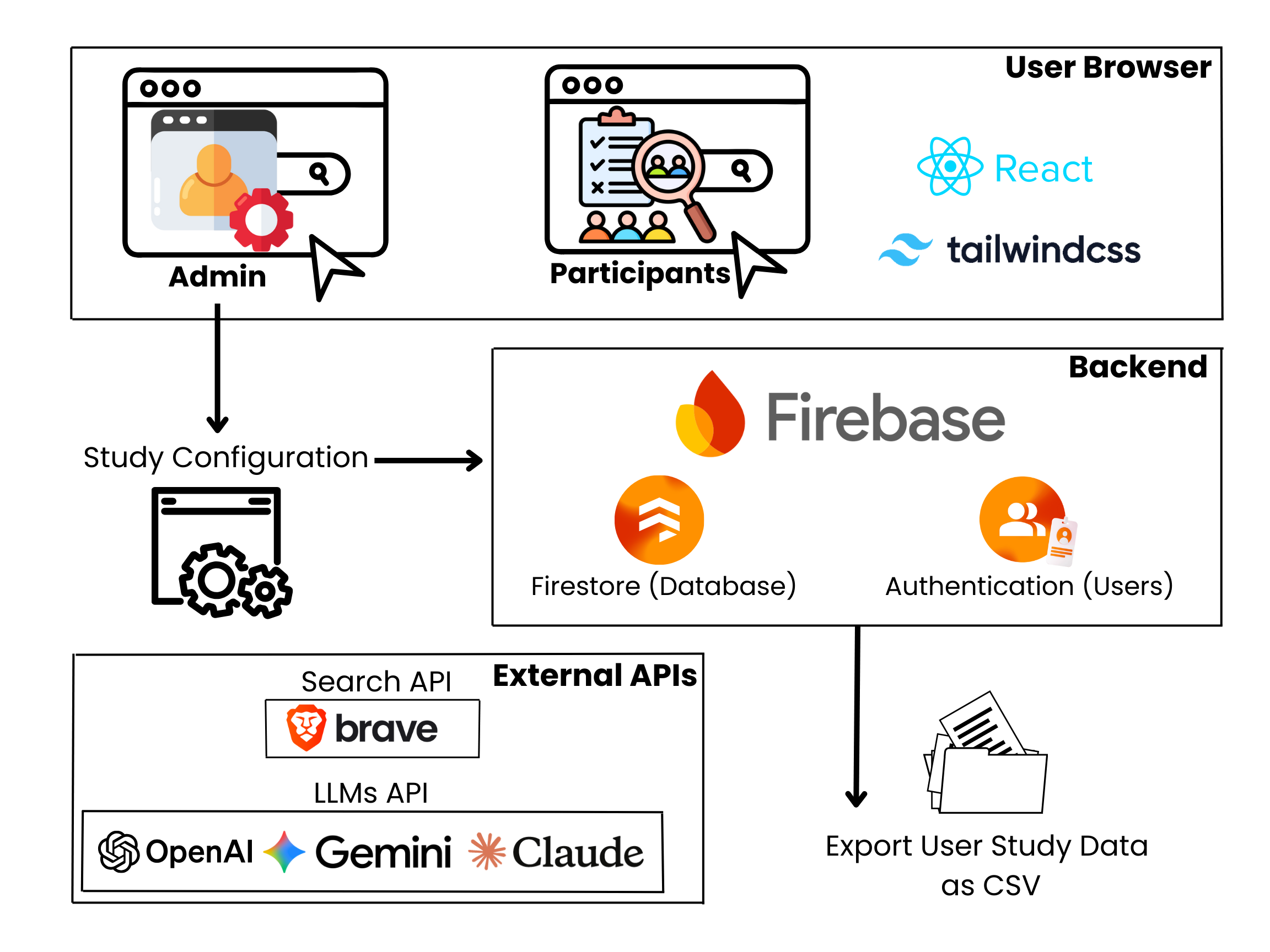}
  \caption{System architecture of ECHO. Administrators configure studies through a Web interface. All data is stored in Firebase and exportable as CSV files for further analysis.}
  \label{fig:architecture}
\end{figure}

\subsubsection{Overview}
ECHO employs a serverless architecture with three layers (Figure~\ref{fig:architecture}). The \textbf{frontend layer} consists of two React-based web applications built with TailwindCSS: an administrator dashboard for study configuration and a participant interface for completing research tasks. The \textbf{backend layer} uses Firebase services, including Firestore for real-time data storage and Firebase Authentication for user management. The \textbf{external API layer} integrates with LLM providers (OpenAI, Google Gemini, Anthropic Claude) for chat tasks and  Search API for web search tasks.

\subsubsection{Data Flow}
Administrators configure study parameters (surveys, task order, API settings) through the dashboard, which stores configurations in Firestore. Participants authenticate via Firebase Authentication and complete assigned tasks. During chat tasks, prompts are sent to the configured LLM API, and responses are streamed back to the interface. During search tasks, queries are sent to the Search API, and results are displayed for participant interaction. All interactions, including timestamps, clicks, and survey responses, are logged to Firestore in real-time. Researchers can export complete study data as structured CSV files directly from the administrator dashboard.

\subsubsection{Interaction Logging}
ECHO captures comprehensive interaction data for both task types:

\textbf{Chat Interactions:} Each prompt-response pair is logged with unique identifiers, full text content, typing timestamps (start and end), and user ratings of chat turns and trajectories.

\textbf{Search Interactions:} Each query logs the search text, typing timestamps, total count of result lists, clicked result URLs with timestamps, and result rankings for clicked URLs items.

\begin{figure}[H]
  \centering
  \includegraphics[width=\columnwidth]{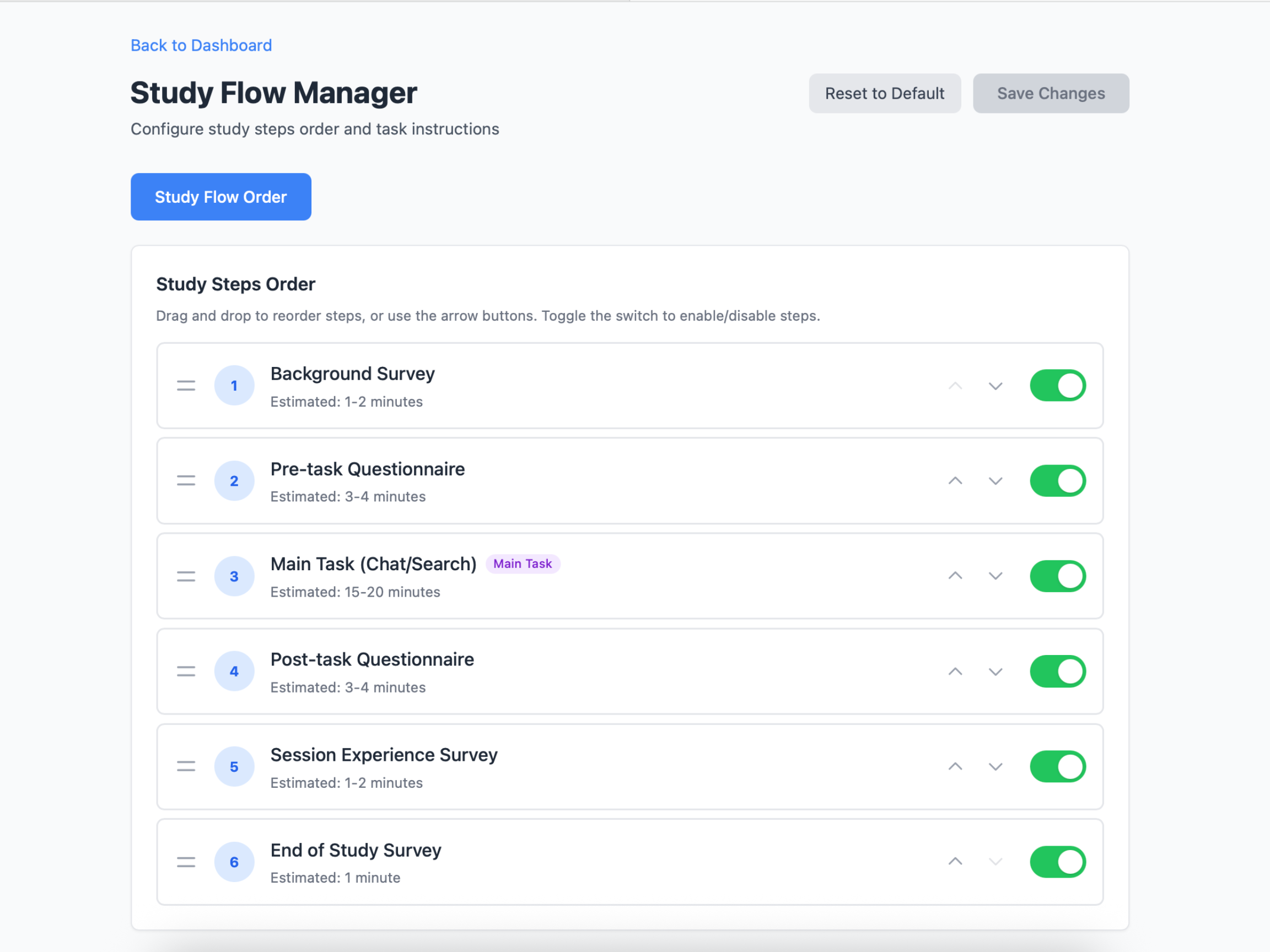}
  \caption{The study flow management interface. Workflow steps can be turned on/off and customized individually and reordered through drag-and-drop interaction.}
  \label{fig:study-flow}
\end{figure}

\subsubsection{Study Flow Management}
The study flow is fully configurable through the administrator dashboard. Default steps include: (1) background survey, (2) pre-task questionnaire, (3) main task (chat or search), (4) post-task questionnaire, (5) experience survey, and (6) end-of-study survey. Administrators can enable, disable, or reorder steps and configure step-specific reminders or attention checks.
\section{Key Features}
ECHO offers a wide variety of key features, which can be easily customized to fit the diverse research needs of different disciplines. 
\subsection{Dual-Modality Support}
ECHO uniquely supports both conversational AI (via LLM APIs) and traditional Web search (via search APIs) within a single platform, enabling direct comparison studies of user behavior across tasks and online information access paradigms.

\subsection{Flexible API Integration}
While defaulting to OpenAI and Brave Search, ECHO's modular design supports integration with alternative providers including Anthropic Claude, Google Gemini, and various search APIs, allowing researchers to study different system configurations.

\subsection{Multi-Mode Survey Configuration}
ECHO provides four editing modes for all survey instruments through the administrator dashboard. \textbf{View Mode} displays current questions with configuration details. \textbf{Reorder Mode} enables drag-and-drop reorganization. \textbf{Form Mode} provides an interactive interface for creating questions with multiple response types (Likert scales, multiple choice, open-ended). \textbf{JSON Mode} allows direct import and export of question configurations. 

\subsection{In-Situ Survey System}
ECHO supports dynamic popup surveys triggered by participant actions. Triggers include: after N prompts submitted, after N responses received, after N search queries, at periodic intervals, or before task submission. This enables researchers to capture contextual feedback without disrupting task flow.

\subsection{Comprehensive Data Export}
ECHO exports structured datasets including: participant registration data, demographic responses, pre/post-task questionnaire responses, complete chat histories with timestamps, search queries with click data, in-situ survey responses, and participant notes.

\section{Potential Research Applications}
The ECHO system can potentially be applied and extended to investigate a wide range of information seeking, evaluation, and human–AI interaction problems across varying computing (both within and beyond SIGIR community), social sciences, and humanities fields. Within IR and HCI, ECHO enables studies that move beyond query- or turn-level evaluation toward longitudinal analyses of interaction behavior and user experience, while requiring little to no custom system implementation. Researchers can examine how users adapt their strategies across multi-turn chat sessions, how expectations and trust evolve when interacting with conversational AI versus traditional Web search, and how different system configurations influence perceived usefulness, satisfaction, task success, and learning progress \cite{yu2025chat}. By \textbf{allowing study workflows, tasks, and questionnaires to be configured through an administrative interface rather than code}, ECHO lowers the barrier to conducting controlled, reproducible studies that combine fine-grained interaction traces with measures of intent fulfillment.

\begin{figure}[h]
  \centering
  \includegraphics[width=\columnwidth]{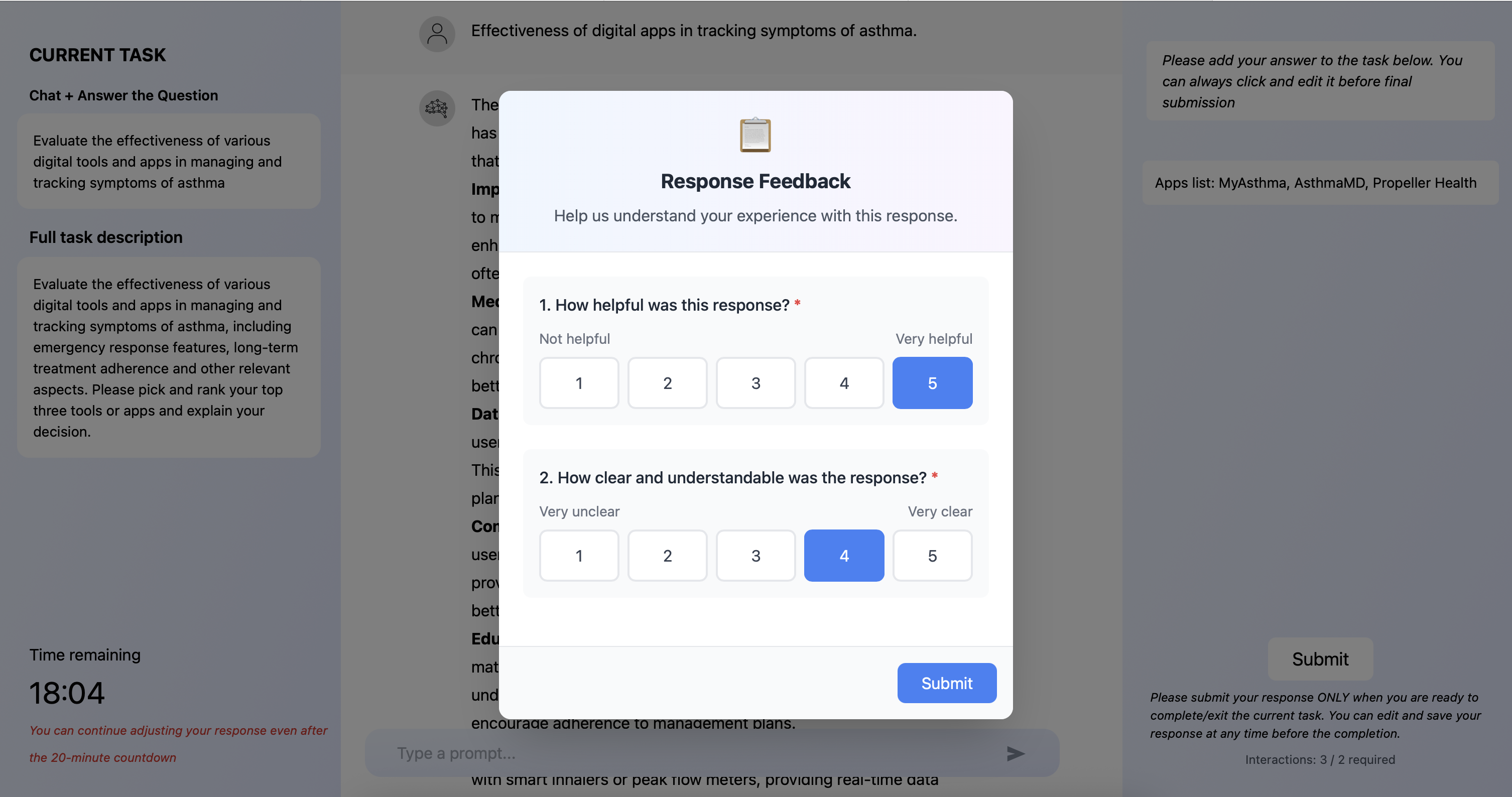}
  \caption{In-situ survey popup triggered during interaction.}
  \label{fig:in-situ}
\end{figure}

Moreover, ECHO provides an accessible experimental environment for behavioral science, social science, and humanities research that seeks to examine how AI-mediated information access shapes judgment, beliefs, and sense-making~\cite{bail2024can, meng2024ai, zhong2024opportunities}. The platform enables researchers with limited programming experience to deploy interactive studies that capture both process-level interaction data and outcome-level measures, such as confidence, perceived credibility, or attitude change. In behavioral research, ECHO can support investigations of cognitive biases, risk perception, or confidence calibration under different conversational or informational framings, with in-situ surveys enabling the measurement of transient mental states during interaction. In social science and humanities contexts, ECHO can be used to study engagement with contested or interpretive content, narrative construction, or perspective-taking, without requiring custom interface development or logging pipelines. 

\section{Conclusion}
Interactive and generative information systems challenge core assumptions of traditional IR evaluation by shifting interaction from isolated queries toward longitudinal, multi turn engagements that shape users’ cognitive states and loads, trust, and decision making~\cite{markwald2023constructing}. Existing tools and experimental systems are not well suited to capture these dynamics or to connect human outcomes to underlying system behavior. ECHO addresses this gap by providing a unified, configurable research platform that supports reproducible, low-code evaluation across both conversational AI and Web search. By integrating fine grained interaction logging, flexible pre task, post task, and in situ measurement, and dual modality task support, ECHO enables researchers to examine trajectory level behavior, expectation shifts, confidence calibration, and outcome formation under realistic conditions. As shared evaluation infrastructure, ECHO lowers technical barriers, promotes methodological consistency, and enables cumulative, interdisciplinary research on human centered evaluation of modern information access systems.

\balance
\bibliographystyle{ACM-Reference-Format}


\begin{thebibliography}{22}


\ifx \showCODEN    \undefined \def \showCODEN     #1{\unskip}     \fi
\ifx \showISBNx    \undefined \def \showISBNx     #1{\unskip}     \fi
\ifx \showISBNxiii \undefined \def \showISBNxiii  #1{\unskip}     \fi
\ifx \showISSN     \undefined \def \showISSN      #1{\unskip}     \fi
\ifx \showLCCN     \undefined \def \showLCCN      #1{\unskip}     \fi
\ifx \shownote     \undefined \def \shownote      #1{#1}          \fi
\ifx \showarticletitle \undefined \def \showarticletitle #1{#1}   \fi
\ifx \showURL      \undefined \def \showURL       {\relax}        \fi
\providecommand\bibfield[2]{#2}
\providecommand\bibinfo[2]{#2}
\providecommand\natexlab[1]{#1}
\providecommand\showeprint[2][]{arXiv:#2}

\bibitem[Azzopardi(2021)]%
        {azzopardi2021cognitive}
\bibfield{author}{\bibinfo{person}{Leif Azzopardi}.} \bibinfo{year}{2021}\natexlab{}.
\newblock \showarticletitle{Cognitive biases in search: a review and reflection of cognitive biases in Information Retrieval}. In \bibinfo{booktitle}{\emph{Proceedings of the 2021 conference on human information interaction and retrieval}}. \bibinfo{pages}{27--37}.
\newblock


\bibitem[Bail(2024)]%
        {bail2024can}
\bibfield{author}{\bibinfo{person}{Christopher~A Bail}.} \bibinfo{year}{2024}\natexlab{}.
\newblock \showarticletitle{Can Generative AI improve social science?}
\newblock \bibinfo{journal}{\emph{Proceedings of the National Academy of Sciences}} \bibinfo{volume}{121}, \bibinfo{number}{21} (\bibinfo{year}{2024}), \bibinfo{pages}{e2314021121}.
\newblock


\bibitem[Carterette et~al\mbox{.}(2016)]%
        {carterette2016evaluating}
\bibfield{author}{\bibinfo{person}{Ben Carterette}, \bibinfo{person}{Paul Clough}, \bibinfo{person}{Mark Hall}, \bibinfo{person}{Evangelos Kanoulas}, {and} \bibinfo{person}{Mark Sanderson}.} \bibinfo{year}{2016}\natexlab{}.
\newblock \showarticletitle{Evaluating retrieval over sessions: The TREC session track 2011-2014}. In \bibinfo{booktitle}{\emph{Proceedings of the 39th International ACM SIGIR conference on Research and Development in Information Retrieval}}. \bibinfo{pages}{685--688}.
\newblock


\bibitem[Chapelle et~al\mbox{.}(2009)]%
        {chapelle2009expected}
\bibfield{author}{\bibinfo{person}{Olivier Chapelle}, \bibinfo{person}{Donald Metlzer}, \bibinfo{person}{Ya Zhang}, {and} \bibinfo{person}{Pierre Grinspan}.} \bibinfo{year}{2009}\natexlab{}.
\newblock \showarticletitle{Expected reciprocal rank for graded relevance}. In \bibinfo{booktitle}{\emph{Proceedings of the 18th ACM conference on Information and knowledge management}}. \bibinfo{pages}{621--630}.
\newblock


\bibitem[Jiang et~al\mbox{.}(2025)]%
        {jiang2025landscape}
\bibfield{author}{\bibinfo{person}{Tianji Jiang}, \bibinfo{person}{Wenqi Li}, {and} \bibinfo{person}{Jiqun Liu}.} \bibinfo{year}{2025}\natexlab{}.
\newblock \showarticletitle{The landscape of data reuse in interactive information retrieval: Motivations, sources, and evaluation of reusability}.
\newblock \bibinfo{journal}{\emph{Journal of the Association for Information Science and Technology}} \bibinfo{volume}{76}, \bibinfo{number}{9} (\bibinfo{year}{2025}), \bibinfo{pages}{1258--1276}.
\newblock


\bibitem[Kelly et~al\mbox{.}(2009)]%
        {kelly2009methods}
\bibfield{author}{\bibinfo{person}{Diane Kelly} {et~al\mbox{.}}} \bibinfo{year}{2009}\natexlab{}.
\newblock \showarticletitle{Methods for evaluating interactive information retrieval systems with users}.
\newblock \bibinfo{journal}{\emph{Foundations and Trends{\textregistered} in Information Retrieval}} \bibinfo{volume}{3}, \bibinfo{number}{1--2} (\bibinfo{year}{2009}), \bibinfo{pages}{1--224}.
\newblock


\bibitem[Lei et~al\mbox{.}(2020)]%
        {lei2020conversational}
\bibfield{author}{\bibinfo{person}{Wenqiang Lei}, \bibinfo{person}{Xiangnan He}, \bibinfo{person}{Maarten de Rijke}, {and} \bibinfo{person}{Tat-Seng Chua}.} \bibinfo{year}{2020}\natexlab{}.
\newblock \showarticletitle{Conversational recommendation: Formulation, methods, and evaluation}. In \bibinfo{booktitle}{\emph{Proceedings of the 43rd International ACM SIGIR Conference on Research and Development in Information Retrieval}}. \bibinfo{pages}{2425--2428}.
\newblock


\bibitem[Liu(2022)]%
        {liu2022toward}
\bibfield{author}{\bibinfo{person}{Jiqun Liu}.} \bibinfo{year}{2022}\natexlab{}.
\newblock \showarticletitle{Toward Cranfield-inspired reusability assessment in interactive information retrieval evaluation}.
\newblock \bibinfo{journal}{\emph{Information Processing \& Management}} \bibinfo{volume}{59}, \bibinfo{number}{5} (\bibinfo{year}{2022}), \bibinfo{pages}{103007}.
\newblock


\bibitem[Liu(2023)]%
        {liu2023toward}
\bibfield{author}{\bibinfo{person}{Jiqun Liu}.} \bibinfo{year}{2023}\natexlab{}.
\newblock \showarticletitle{Toward a two-sided fairness framework in search and recommendation}. In \bibinfo{booktitle}{\emph{Proceedings of the 2023 Conference on Human Information Interaction and Retrieval}}. \bibinfo{pages}{236--246}.
\newblock


\bibitem[Liu and Azzopardi(2025)]%
        {liu2025report}
\bibfield{author}{\bibinfo{person}{Jiqun Liu} {and} \bibinfo{person}{Leif Azzopardi}.} \bibinfo{year}{2025}\natexlab{}.
\newblock \showarticletitle{Report on the CHIIR 2024, SIGIR 2024, and SIGIR-AP 2024 tutorials on Characterizing, Evaluating, and Mitigating Cognitive Biases in Interactive and Generative IR}. In \bibinfo{booktitle}{\emph{ACM SIGIR Forum}}, Vol.~\bibinfo{volume}{59}. ACM New York, NY, USA, \bibinfo{pages}{1--13}.
\newblock


\bibitem[Liu and Shah(2019)]%
        {liu2019interactive}
\bibfield{author}{\bibinfo{person}{Jiqun Liu} {and} \bibinfo{person}{Chirag Shah}.} \bibinfo{year}{2019}\natexlab{}.
\newblock \bibinfo{booktitle}{\emph{Interactive IR user study design, evaluation, and reporting}}.
\newblock \bibinfo{publisher}{Morgan \& Claypool Publishers}.
\newblock


\bibitem[Markwald et~al\mbox{.}(2023)]%
        {markwald2023constructing}
\bibfield{author}{\bibinfo{person}{Marco Markwald}, \bibinfo{person}{Jiqun Liu}, {and} \bibinfo{person}{Ran Yu}.} \bibinfo{year}{2023}\natexlab{}.
\newblock \showarticletitle{Constructing and meta-evaluating state-aware evaluation metrics for interactive search systems}.
\newblock \bibinfo{journal}{\emph{Information Retrieval Journal}} \bibinfo{volume}{26}, \bibinfo{number}{1} (\bibinfo{year}{2023}), \bibinfo{pages}{10}.
\newblock


\bibitem[Meng et~al\mbox{.}(2026)]%
        {meng2026re}
\bibfield{author}{\bibinfo{person}{Chuan Meng}, \bibinfo{person}{Jiqun Liu}, \bibinfo{person}{Mohammad Aliannejadi}, \bibinfo{person}{Fengran Mo}, \bibinfo{person}{Jeff Dalton}, {and} \bibinfo{person}{Maarten de Rijke}.} \bibinfo{year}{2026}\natexlab{}.
\newblock \showarticletitle{Re-Rankers as Relevance Judges}.
\newblock \bibinfo{journal}{\emph{arXiv preprint arXiv:2601.04455}} (\bibinfo{year}{2026}).
\newblock


\bibitem[Meng(2024)]%
        {meng2024ai}
\bibfield{author}{\bibinfo{person}{Juanjuan Meng}.} \bibinfo{year}{2024}\natexlab{}.
\newblock \showarticletitle{AI emerges as the frontier in behavioral science}.
\newblock \bibinfo{journal}{\emph{Proceedings of the National Academy of Sciences}} \bibinfo{volume}{121}, \bibinfo{number}{10} (\bibinfo{year}{2024}), \bibinfo{pages}{e2401336121}.
\newblock


\bibitem[Mo et~al\mbox{.}(2025)]%
        {mo2025survey}
\bibfield{author}{\bibinfo{person}{Fengran Mo}, \bibinfo{person}{Kelong Mao}, \bibinfo{person}{Ziliang Zhao}, \bibinfo{person}{Hongjin Qian}, \bibinfo{person}{Haonan Chen}, \bibinfo{person}{Yiruo Cheng}, \bibinfo{person}{Xiaoxi Li}, \bibinfo{person}{Yutao Zhu}, \bibinfo{person}{Zhicheng Dou}, {and} \bibinfo{person}{Jian-Yun Nie}.} \bibinfo{year}{2025}\natexlab{}.
\newblock \showarticletitle{A survey of conversational search}.
\newblock \bibinfo{journal}{\emph{ACM Transactions on Information Systems}} \bibinfo{volume}{43}, \bibinfo{number}{6} (\bibinfo{year}{2025}), \bibinfo{pages}{1--50}.
\newblock


\bibitem[Shah et~al\mbox{.}(2023)]%
        {shah2023taking}
\bibfield{author}{\bibinfo{person}{Chirag Shah}, \bibinfo{person}{Ryen White}, \bibinfo{person}{Paul Thomas}, \bibinfo{person}{Bhaskar Mitra}, \bibinfo{person}{Shawon Sarkar}, {and} \bibinfo{person}{Nicholas Belkin}.} \bibinfo{year}{2023}\natexlab{}.
\newblock \showarticletitle{Taking search to task}. In \bibinfo{booktitle}{\emph{Proceedings of the 2023 Conference on Human Information Interaction and Retrieval}}. \bibinfo{pages}{1--13}.
\newblock


\bibitem[Voorhees et~al\mbox{.}(2005)]%
        {voorhees2005trec}
\bibfield{author}{\bibinfo{person}{Ellen~M Voorhees}, \bibinfo{person}{Donna~K Harman}, {et~al\mbox{.}}} \bibinfo{year}{2005}\natexlab{}.
\newblock \bibinfo{booktitle}{\emph{TREC: Experiment and evaluation in information retrieval}}. Vol.~\bibinfo{volume}{63}.
\newblock \bibinfo{publisher}{MIT press Cambridge}.
\newblock


\bibitem[Wan et~al\mbox{.}(2023)]%
        {wan2023biasasker}
\bibfield{author}{\bibinfo{person}{Yuxuan Wan}, \bibinfo{person}{Wenxuan Wang}, \bibinfo{person}{Pinjia He}, \bibinfo{person}{Jiazhen Gu}, \bibinfo{person}{Haonan Bai}, {and} \bibinfo{person}{Michael~R Lyu}.} \bibinfo{year}{2023}\natexlab{}.
\newblock \showarticletitle{Biasasker: Measuring the bias in conversational ai system}. In \bibinfo{booktitle}{\emph{Proceedings of the 31st ACM Joint European Software Engineering Conference and Symposium on the Foundations of Software Engineering}}. \bibinfo{pages}{515--527}.
\newblock


\bibitem[White and Shah(2025)]%
        {white2025information}
\bibfield{author}{\bibinfo{person}{Ryen~W White} {and} \bibinfo{person}{Chirag Shah}.} \bibinfo{year}{2025}\natexlab{}.
\newblock \bibinfo{booktitle}{\emph{Information Access in the Era of Generative AI}}.
\newblock \bibinfo{publisher}{Springer}.
\newblock


\bibitem[Yu and Liu(2025)]%
        {yu2025chat}
\bibfield{author}{\bibinfo{person}{Ran Yu} {and} \bibinfo{person}{Jiqun Liu}.} \bibinfo{year}{2025}\natexlab{}.
\newblock \showarticletitle{Chat as Learning in Interactive Information Retrieval and Generation}. In \bibinfo{booktitle}{\emph{ACM SIGIR Forum}}, Vol.~\bibinfo{volume}{59}. ACM New York, NY, USA, \bibinfo{pages}{1--19}.
\newblock


\bibitem[Zamani et~al\mbox{.}(2023)]%
        {zamani2023conversational}
\bibfield{author}{\bibinfo{person}{Hamed Zamani}, \bibinfo{person}{Johanne~R Trippas}, \bibinfo{person}{Jeff Dalton}, \bibinfo{person}{Filip Radlinski}, {et~al\mbox{.}}} \bibinfo{year}{2023}\natexlab{}.
\newblock \showarticletitle{Conversational information seeking}.
\newblock \bibinfo{journal}{\emph{Foundations and Trends{\textregistered} in Information Retrieval}} \bibinfo{volume}{17}, \bibinfo{number}{3-4} (\bibinfo{year}{2023}), \bibinfo{pages}{244--456}.
\newblock


\bibitem[Zhong et~al\mbox{.}(2024)]%
        {zhong2024opportunities}
\bibfield{author}{\bibinfo{person}{Tianyang Zhong}, \bibinfo{person}{Zhenyuan Yang}, \bibinfo{person}{Zhengliang Liu}, \bibinfo{person}{Ruidong Zhang}, \bibinfo{person}{Weihang You}, \bibinfo{person}{Yiheng Liu}, \bibinfo{person}{Haiyang Sun}, \bibinfo{person}{Yi Pan}, \bibinfo{person}{Yiwei Li}, \bibinfo{person}{Yifan Zhou}, {et~al\mbox{.}}} \bibinfo{year}{2024}\natexlab{}.
\newblock \showarticletitle{Opportunities and challenges of large language models for low-resource languages in humanities research}.
\newblock \bibinfo{journal}{\emph{arXiv preprint arXiv:2412.04497}} (\bibinfo{year}{2024}).
\newblock


\end{thebibliography}

\end{document}